\documentclass[aip,apl,graphicx,amsmath,amssymb]{revtex4-1}

\usepackage[utf8]{inputenc} % allow utf-8 input
\usepackage{hyperref}       % hyperlinks
\usepackage{url}            % simple URL typesetting
\usepackage{booktabs}       % professional-quality tables
\usepackage{amsfonts}       % blackboard math symbols
\usepackage{graphicx}       % graphics
\usepackage[nameinlink]{cleveref}
\usepackage{cprotect}
\usepackage{color}

\draft % marks overfull lines with a black rule on the right

\begin{document}

%\preprint{}

\title{Multi-resolution Physics-Aware Recurrent Convolutional Neural Network for Complex Flows} %Title of paper



\author{Xinlun Cheng}
\affiliation{School of Data Science, University of Virginia, 1919 Ivy
Rd., Charlottesville, 22903, VA, USA}

\author{Joseph Choi}
\affiliation{School of Data Science, University of Virginia, 1919 Ivy
Rd., Charlottesville, 22903, VA, USA}

\author{H.S. Udaykumar}
\affiliation{Department of Mechanical Engineering, University of Iowa, 103 S. Capitol St., Iowa City, 52242, IA, USA}

\author{Stephen Baek}
\email{mwn4yc@virginia.edu}
\affiliation{School of Data Science, University of Virginia, 1919 Ivy Rd., Charlottesville, 22903, VA, USA}
\affiliation{Department of Mechanical and Aerospace Engineering, University of Virginia, 351 McCormick Rd., Charlottesville, 22904, VA, USA}
%\thanks{}


\date{\today}

\begin{abstract}
We present MRPARCv2, Multi-resolution Physics-Aware Recurrent Convolutional Neural Network, designed to model complex flows by embedding the structure of advection-diffusion-reaction equations and leveraging a multi-resolution architecture. MRPARCv2 introduces hierarchical discretization and cross-resolution feature communication to improve accuracy and efficiency of flow simulations. We evaluate the model on a challenging 2D turbulent radiative layer dataset from The Well multi-physics benchmark repository and demonstrate significant improvements when compared to the single resolution baseline model, in both Variance Scaled Root Mean Squared Error (VRMSE) and physics-driven metrics, including turbulent kinetic energy spectra and mass–temperature distributions. Despite having 30\% less trainable parameters, MRPARCv2 outperforms its predecessor by up to 50\% in roll-out prediction error and 86\% in spectral error. A preliminary study on uncertainty quantification was performed, and we also analyze the model’s performance under different levels of abstractions of the flow, specifically on sampling subsets of field variables. We find that the absence of physical constraints on equation of state (EOS) in the network architecture leads to degraded accuracy. A variable substitution experiment confirms that this issue persists regardless of which physical quantity is predicted directly. Our findings highlight the advantages of multi-resolution inductive bias for capturing multi-scale flow dynamics and suggest the need for future PIML models to embed EOS knowledge to enhance physical fidelity.
\end{abstract}

\pacs{}% insert suggested PACS numbers in braces on next line

\maketitle %\maketitle must follow title, authors, abstract and \pacs

\textbf{Topics:}

Deep learning, Machine learning, Physics-informed machine learning, Fluid mechanics

\section{Introduction}
With the recent advancements in artificial intelligence (AI) and machine learning (ML), physics-informed machine learning (PIML) has become a widely adopted technique for addressing the challenges of simulating flow fields with complex patterns. It provides a powerful framework for predicting flow field evolution governed by physical laws, reducing the computational costs associated with traditional numerical methods. Moreover, it alleviates the reliance on the large volumes of training data typically required by physics-agnostic machine learning approaches. This capability is achieved in PIML by integrating domain knowledge, typically expressed as partial differential equations (PDEs), directly into the machine learning process\citep{karniadakis2021physics}. Embedding the governing physical equations of a system into the neural network’s architecture (inductive bias\citep{pateras2023taxonomic}), or more commonly into its loss function or optimization process (learning bias\citep{pateras2023taxonomic}), ensures that the neural network is trained not only to fit available training data , but also to remain consistent with the governing physical laws.

This paradigm offers several advantages for the computation of complex fluid flows. Firstly, PIML substantially reduces the dependence on extensive datasets, as the physical laws provide prior knowledge that guides the learning process\citep{karniadakis2021physics}. This advantage is particularly important when the acquisition of high-fidelity training data is computationally expensive or experimentally infeasible. Secondly, by enforcing physical consistency, PIML models demonstrate enhanced robustness and generalizability. This property ensures that predictions remain physically plausible and reliable, even in regions with limited data coverage or under previously unseen operating conditions\citep{stiasny2021learning,meng2025physics}. Lastly, PIML enables new opportunities for design optimization by markedly accelerating the computation of quantities of interest from specified initial conditions. Such tasks were historically computationally prohibitive due to the overhead of conventional numerical simulations\citep{choi2023artificial}.

For instance, physics-informed neural networks (PINNs\citep{raissi2019physics}) are among the most widely used approaches in the computational fluid dynamics (CFD) community. By directly embedding the governing PDEs into the loss function (training objective), this approach guides the neural network to satisfy both observed data and physical constraints. This enables effective learning even when only limited training samples are available. Recent advancements in PINNs for complex flows include applications to adverse pressure gradient (APG) turbulent boundary layers\citep{hanrahan2023studying}, flow field reconstruction from RANS simulations\citep{patel2024turbulence}, and composite porous-fluid systems\citep{jang2024physics}. However, PINN models often exhibit limited generalization to unseen initial or boundary conditions, even when the governing equations remain unchanged. This shortcoming stems from their fundamental design\citep{wang2024advancing,zhang2025combining}. PINNs attempt to fit highly nonlinear functions that map space-time coordinates to solutions of the governing equations. They do so by minimizing a combined loss consisting of the equation residuals and the data discrepancy.

Another prominent class of PIML models is neural operators (NOs\citep{kovachki2023neural}), with the Fourier Neural Operator (FNO\citep{li2020fourier}) being the most widely adopted variant. NOs are deep learning architectures designed to learn mappings between infinite-dimensional function spaces, and they directly approximate solution operators of PDEs. This mathematical formulation enables NOs to generalize across different initial conditions and grid resolutions without the need for retraining. NOs have been successfully applied to 3D incompressible turbulent flows\citep{zhao2025lesnets}, Rayleigh-Benard convection\citep{rahman2024pretraining}, and weather forecasting\citep{kurth2023fourcastnet}. However, they are known to suffer performance degradation as system complexity increases. In addition, they exhibit spectral bias, which often leads to over-smoothing and the loss of high-frequency content. These features are critical for accurately capturing complex flow patterns, particularly in long-term predictions\citep{zheng2025cf}.

In recent years, there has been growing interest in inductive bias PIML models\citep{meng2025physics}. Unlike PINNs and NOs, where governing equations are typically incorporated into the training process via the loss function, inductive bias models aim to embed physics constraints (e.g., symmetries, conservation laws, and the structure of governing equations) directly into the neural network architecture. For instance, Hamiltonian neural networks encode energy conservation laws into their architecture. They have demonstrated superior stability and accuracy in modeling nonlinear oscillators and the Henon-Heiles system\citep{mattheakis2022hamiltonian}. The Physics-Aware Recurrent Convolutional Neural Network (PARC\citep{nguyen2024parcv2}), which embeds the structure of advection-diffusion-reaction (ADR) equations into the network design, has achieved state-of-the-art accuracy in several applications. These include modeling the initiation of energetic materials under strong shock, material shear accumulation under weak shock (Cheng et al. 2025, in press), predicting supersonic fluid flow under varying conditions\citep{cheng2024physics}, and optimizing ventilation performance in confined spaces\citep{kim2025physics}. Equivariant neural networks have also been successfully applied to atomic and molecular dynamics simulations\citep{batzner20223}. In addition, they have been used in the computation of electron densities in nitrogen molecules and hydrogen chains\citep{pfau2020ab}, as well as in medical image analysis\citep{bekkers2018roto}. \citet{fukami2023grasping} proposed observable-augmented manifold learning, in which auto-encoders are trained to reconstruct flow fields while simultaneously predicting physical observables (e.g., lift coefficients). This approach achieves efficient dimensionality reduction and simplifies latent dynamics, enabling accurate modeling of gust–wing interactions and rough-wall turbulent drag \citep{nair2025rough}. The resulting low-dimensional representations support applications in vehicle geometry optimization \citep{tran2024aerodynamics}, flow reconstruction from sparse measurements \citep{mousavi2025low}, and control of transient lift under extreme perturbations\citep{fukami2024data}.

As the archetype of spatio-temporally varying and evolving complex flows, turbulent flows have been studied extensively across a wide range of natural science and engineering applications\citep{pope2001turbulent}. Turbulent flows are characterized by chaotic and irregular fluctuations in velocity, pressure, and other fluid properties. They exhibit non-parallel streamlines, high rates of lateral mixing, and a pervasive presence of eddies that vary substantially in both size and direction\citep{launder1983numerical}. This intrinsic randomness and multiscale complexity render them exceptionally difficult to describe and to model mathematically\citep{kajishima2016computational}.

Directly solving the governing equations that encapsulate the fundamental principles of energy, momentum, and mass conservation via Direct Numerical Simulation (DNS) often incurs prohibitively high computational costs. This arises from the need to resolve the extremely fine spatial and temporal scales inherent to turbulent flows\citep{argyropoulos2015recent}. Consequently, for many real-world turbulent flow scenarios, simulations rely on turbulence models such as Reynolds-Averaged Navier–Stokes (RANS\citep{reynolds1895iv,yusuf2020short}) or Large Eddy Simulation (LES\citep{smagorinsky1963general}). However, RANS models exhibit shortcomings when applied to complex engineering flows. These include cases with strong anisotropy, significant streamline curvature, flow separation, or non-equilibrium states\citep{speziale1998turbulence,girimaji2006partially,menter2009review,spalart2015philosophies}. Although LES provides higher fidelity than RANS and captures more flow physics, it still demands substantial computational resources. Moreover, the subgrid-scale models used in LES may fail to represent the full range of turbulent interactions\citep{meneveau2000scale,georgiadis2010large,piomelli2014large}. Other numerical methods, such as hybrid RANS–LES and Detached Eddy Simulation (DES), have gained popularity due to their ability to balance computational workload and accuracy\citep{frohlich2008hybrid,spalart2009detached,deck2012recent,chaouat2017state}. Nevertheless, a persistent challenge remains: high-fidelity methods are computationally prohibitive, while practical models retain inherent accuracy limitations. This creates a fundamental gap in the ability to capture the diverse flow phenomena observed across scales and conditions in experiments.

In this work, we present the Multi-resolution Physics-Aware Recurrent Convolutional Neural Network (MRPARCv2), a neural network that integrates the structure of ADR equations, a multi-resolution approach, and hard boundary condition enforcement for modeling complex fluid flows. In \Cref{sec:architecture}, we describe the architecture of our neural network. In \Cref{sec:data}, we introduce the turbulent radiative layer problem and outline the training strategy for our machine learning model. In \Cref{sec:results}, we compare MRPARCv2 with the newest
variant of PARC, PARCv2\citep{nguyen2024parcv2}, in terms of both prediction error and physics-driven metrics, demonstrating the superior accuracy of MRPARCv2. In \Cref{sec:discussion}, we conduct preliminary study on uncertainty quantification, examine the network's ability to learn the governing equations, discuss current limitations, and suggest directions for future development. Finally, in \Cref{sec:conclusion}, we summarize our findings.

\section{Machine Learning Architecture}\label{sec:architecture}
\subsection{Comparison model: baseline single-resolution PARCv2}
We adopted PARCv2\citep{nguyen2024parcv2} as the baseline AI/ML model due to its demonstrated capability in modeling complex transient physical phenomena, such as supersonic flow\citep{cheng2024physics} and strong shock loading in complex material systems with nonlinear reactions\citep{nguyen2024parcv2}. Several improvements have been made to the original architecture:

\begin{itemize}
    \item In the original architecture of PARCV2 in \citet{nguyen2024parcv2} and \citet{cheng2024physics}, the top, bottom, left, and right boundaries were required to use the same boundary conditions. We redesigned all sub-modules to allow different boundary conditions to be applied in each direction. Boundary conditions are enforced strictly and explicitly through different padding scheme, in a manner similar to \citet{ren2022phycrnet}. For this particular problem, periodic boundary condition in X direction is enforced through circular padding, and zero gradient boundary condition in Y direction is enforced through reflection padding.
    \item The original architecture includes multiple copies of the advection and diffusion sub-modules, increasing GPU and CPU memory usage. Since these sub-modules do not contain learnable parameters, we modified the computational graph and these sub-modules to enable reuse.
    \item The original architecture could not handle fields with constant values or fields lacking advection or diffusion terms. We extended the algorithm to support these cases.
\end{itemize}

The baseline model employs a U-Net\citep{ronneberger2015u} feature extractor that processes the input field and generates a latent representation with 64 channels. These learned features are then combined with explicitly computed advection terms to form a unified representation, which is used to predict the time derivatives of the physical fields. The resulting time derivatives are integrated using a fourth-order Runge-Kutta (RK4\citep{runge1895numerische,kutta1901beitrag}) scheme to advance the solution to the next time step. This hybrid approach leverages both data-driven feature extraction and physics-informed computation, enhancing the model’s ability to capture complex spatiotemporal dynamics. The baseline model contains a total of 5.5 million trainable parameters.

\subsection{Multi-resolution PARCv2 (MRPARCv2)}
Many machine learning algorithms have been proposed to exploit the advantages of using multiple resolutions. \citet{ke2017multigrid} investigated the benefits of exchanging information between adjacent resolution levels during convolution operations for image classification. \citet{wang2020multi} proposed the multi-resolution deep convolutional neural network (MCNN), a novel architecture that reconstructs images across a wide range of resolutions and showed that it outperforms the traditional U-Net in image-to-image inverse problems. One major advantage of multi-resolution models is their ability to overcome spectral bias. Although the universal approximation theorem guarantees that a neural network of sufficient depth can approximate any function to arbitrary accuracy, it has been shown that machine learning models tend to learn dominant low-frequency components first, requiring impractically long training times to capture high-frequency signals\citep{xu2019training,rahaman2019spectral,xu2024overview}. However, this issue can be mitigated using multi-resolution models, since high-frequency prediction errors on coarser grids appear as lower-frequency errors on finer grids. Given that turbulent flow is inherently multi-resolution, a multi-discretization approach aligns with the philosophy of inductive bias -- the guiding design principle behind the success of models like PARCv2.

\begin{figure*}
    \centering
    \includegraphics[width=\textwidth]{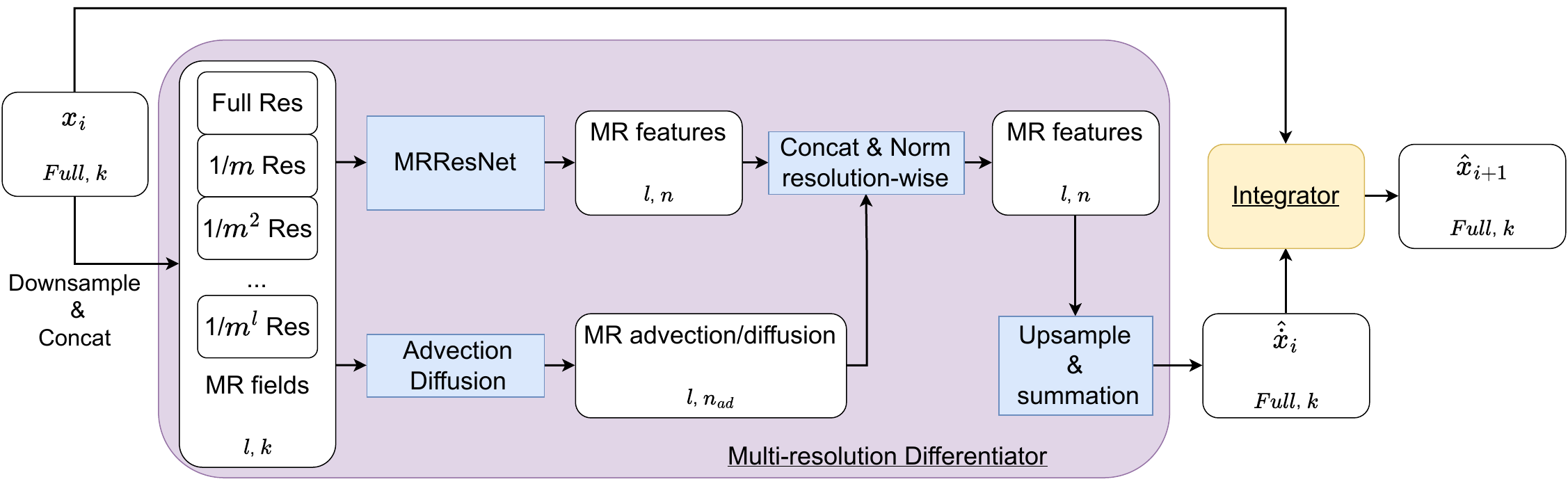}
    \caption{Architecture diagram of MRPARCv2. Standard rectangles denote modules, while rounded rectangles represent inputs, outputs, and intermediate tensors. The number of resolution levels and the number of channels per resolution level are indicated for both module outputs and tensors. Refer to \Cref{fig:modules} for the diagram of each module.}
    \label{fig:architecture}
\end{figure*}

\begin{figure*}
    \centering
    \includegraphics[width=\textwidth]{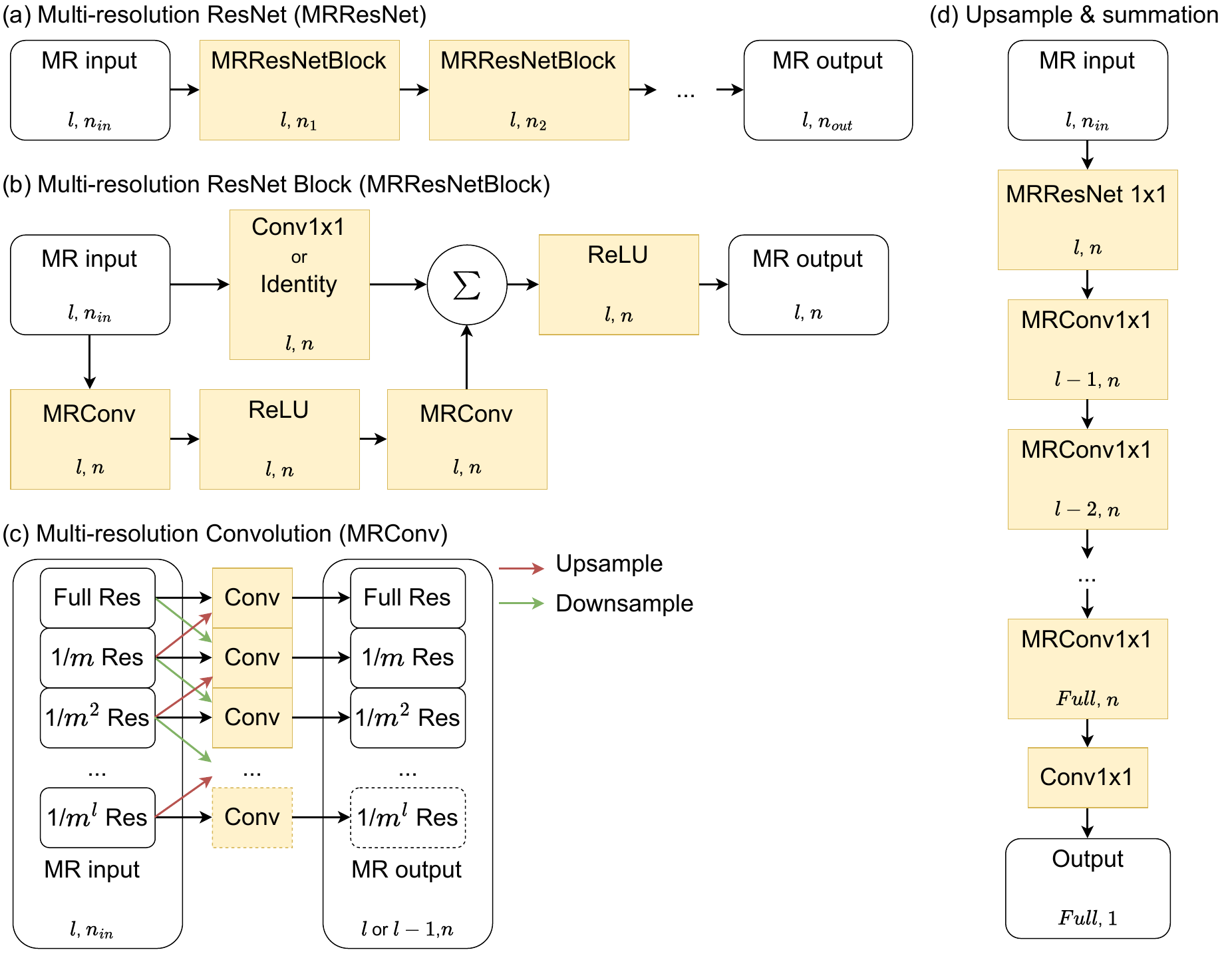}
    \caption{Module diagram of MRPARCv2. Standard rectangles denote modules, while rounded rectangles represent inputs, outputs, and intermediate tensors. The number of resolution levels and the number of channels per resolution level are indicated for both module outputs and tensors.}
    \label{fig:modules}
\end{figure*}

Motivated by this reasoning, we propose and evaluate the performance of Multi-resolution PARCv2 (MRPARCv2). The architecture of our neural network is illustrated in \Cref{fig:architecture}. Given the field values at time step $i$, the model predicts the values at the next time step, $i+1$. The input is downsampled to $l-1$ coarser resolutions using average pooling, reducing the spatial resolution by a factor of $1/m$ at each level. A Multi-Resolution ResNet (\verb|MRResNet|; \Cref{fig:modules}a) then extracts learnable features across the $l$ resolutions (the finest and the $l-1$ downsampled). Each convolutional operation within \verb|MRResNet| (\verb|MRConv|; \Cref{fig:modules}c) allows for information exchange between adjacent finer and coarser grids. At each level, $n$ features are extracted and concatenated with advection terms explicitly computed from the corresponding input resolution (\verb|Concat & Norm resolution-wise| module). These features are then successively interpolated from coarser to finer grids and aggregated at the highest resolution via the \verb|Upsample & Summation| module (\Cref{fig:modules}e), in which a series of \verb|MRConv| sub-modules reduce the number of resolution levels by one each time. The final output is the predicted time derivative for each field, which is then integrated forward in time.

While increasing the number of resolutions ($l$) generally improves model performance, it also incurs higher GPU memory usage. To balance computational cost and accuracy for this task, we use $l=3$ resolution levels and $n=64$ features per level. The MRResNet feature extractor consists of three MRResNetBlocks with $3 \times 3$ convolution kernels. The \verb|Upsample & Summation| module contains two such blocks, and each output channel has its own dedicated module. The time integration method is identical to that used in the baseline PARCv2 model. In total, MRPARCv2 contains 3.9 million trainable parameters, $\sim$30\% less than the baseline single-resolution model and therefore theoretically exhibits less degrees of freedom (dof) than the baseline.

\section{Data and Training Process}\label{sec:data}
We examine the performance of MRPARCv2 on 2D simulations of turbulent radiative layers, which are part of The Well multi-physics dataset\citep{ohana2024well}. In many applications, such as astrophysical systems of interstellar and circumgalactic medium, it is common for hot gas to move relative to cold gas at highly subsonic velocities. Although both the hot and cold phases are in thermal equilibrium, this configuration is unstable to the Kelvin–Helmholtz instability, leading to turbulence and mixing between the two phases\citep{fielding2020multiphase}. Instabilities, followed by turbulence in heterogeneous media occur in many other natural (e.g. ocean currents\citep{smyth2012ocean}, atmospheric science\citep{conrick2018simulated}) and industrial applications (e.g. battery failure modes\citep{gallaway2010lateral}, oil recovery\citep{lake1989enhanced}) as well. The governing equations of the physical system underlying such unstable and spatiotemporally complex flows are as follows:
\begin{equation}
    \frac{\partial\rho}{\partial t} +\nabla\cdot(\rho\vec{v})=0
\end{equation}\label{eqn:mass}
\begin{equation}
    \frac{\partial\rho\vec{v}}{\partial t} +\nabla\cdot(\rho\vec{v}\vec{v}+P)=0
\end{equation}\label{eqn:momentum}
\begin{equation}
    \frac{\partial E}{\partial t} +\nabla\cdot((E+P)\vec{v})=-\frac{E}{t_{cool}}
\end{equation}\label{eqn:energy}
where $\rho$ is the density, $P$ is the pressure, $\vec{v}$ is the velocity vector, $E=\rho e+\frac{1}{2}\rho\vec{v}\cdot\vec{v}$ is the total energy, and $e$ is the internal energy density. The equation of state is ideal gas law, which is
\begin{equation}\label{eqn:eos}
    P=(\gamma-1)\rho e, \gamma=5/3
\end{equation}
For details on the simulation and dataset setup, please refer to \citet{fielding2020multiphase} and \citet{ohana2024well}. We use the same training-validation-test split as in The Well dataset; however, due to limited GPU availability, we exclude all cases with $t_{cool} = 0.03$ and $t_{cool} = 3.16$. The dataset provides four fields: density $\rho$, pressure $P$, X velocity $U$, and Y velocity $V$. We manually appended an additional constant channel representing $t_{cool}$.

Unlike the models presented in \citet{ohana2024well}, which are provided with four consecutive timesteps as input, both the baseline and MRPARCv2 models take only a single timestep as input and predict as many future timesteps as specified by the user. The loss function used is mean absolute error (MAE, L1), and training is performed using the Adam optimizer\citep{kingma2017adam}. No regularization techniques are applied. Each model is first trained to convergence on predicting the next timestep with learning rate $10^{-4}$, and then further trained to predict the subsequent three timesteps with the learning rate $10^{-5}$. All subsequent analyses are performed on the hold-out test set unless otherwise stated in the text or figure captions.

The models were trained on Nvidia A100 GPU using PyTorch\citep{ansel2024pytorch}. The project code are publicly available at \url{https://github.com/chengxinlun/parcv2_trl2d}. MRPARCv2 and baseline PARCv2 modules and sub-modules can be found at \url{https://github.com/chengxinlun/MultiPARC}.

\section{Results}\label{sec:results}
\subsection{Variance-scaled Root Mean Squared Error (VRMSE) and Visual Inspection}
We begin by evaluating the variance-scaled root mean square error (VRMSE) of each field for both MRPARCv2 and the baseline PARCv2 model. For comparison, we also include the best-performing model reported in \citet{ohana2024well}, the ConvNeXt U-Net\footnote{Pretrained weights can be found on Huggingface}. VRMSE is defined as the root mean square error (RMSE) scaled by the spatial variance of the corresponding field. This normalization enables fair comparison of model performance across different channels and is also the recommended evaluation metric in the dataset publication. It is important to note that a model predicting only the spatial mean of a field would yield a VRMSE of 1.

\subsubsection{Single-step prediction}
We first evaluate the single-step prediction VRMSE, as shown in \Cref{tab:vrmse_single}. For each simulation in the test set, the ground truth at timestep $t$ is provided as input to the model, and the prediction for the next timestep $t+1$ is compared to the corresponding ground truth. This process is repeated across all timesteps in the sequence. It is worth noting that, due to its design, the ConvNeXt U-Net model from \citet{ohana2024well} takes four consecutive timesteps as input and contains approximately 18 million trainable parameters, roughly 3.3 times that of baseline PARCv2 and 4.6 times that of MRPARCv2.

\begin{table*}[!ht]
    \centering
    \resizebox{\textwidth}{!}{\begin{tabular}{c|cccc}
    \hline
    Model & VRMSE($\rho$) & VRMSE($P$) & VRMSE($U$) & VRMSE($V$) \\
    \hline
    ConvNext U-Net\citep{ohana2024well} & 0.0645 & 0.4207 & 0.1209 & 0.1783 \\
    Baseline PARCv2 (\citet{nguyen2024parcv2} improved)& 0.0391 & 0.3443 & 0.0900 & 0.1383 \\
    MRPARCv2 (This work) & \textbf{0.0313} & \textbf{0.2787} & \textbf{0.0778} & \textbf{0.1150}\\
    \hline
    \end{tabular}}
    \caption{Single-step prediction VRMSE comparison. Model with smallest VRMSE are in bold for each channel.}
    \label{tab:vrmse_single}
\end{table*}

We observed that MRPARCv2 achieves approximately a $\sim$17\% reduction in VRMSE across all channels, with the largest improvement in density ($\sim$20\%) and the smallest in X velocity ($\sim$14\%). Given that MRPARCv2 has 30\% less trainable parameters than the baseline, this improvement can be attributed to the incorporation of the multi-resolution architecture. Both the baseline and MRPARCv2 models outperform the best-performing model in \citet{ohana2024well}, despite having only one-third and one-fifth the number of trainable parameters, respectively. This could be attributed to the enhanced parameter efficiency of models with strong inductive biases, a desirable trait in data limited regimes. Alternatively, the ML model presented in \citet{ohana2024well} might be significantly over-parameterized. When comparing performance across different fields, the pressure field consistently exhibits the highest VRMSE, regardless of the model used, a trend also reported in \citet{ohana2024well}.

\subsubsection{Roll-out prediction}
In real-world applications of PIML models, ground truth inputs are typically unavailable during inference, requiring the model to recursively predict the full sequence through roll-out. We therefore evaluate the roll-out performance of each model. Due to the design of the ConvNeXt U-Net model from \citet{ohana2024well}, which takes the first four ground truth snapshots at $t = 0, 1, 2, 3$ as input, the roll-out prediction for both PARCv2 and MRPARCv2 is initiated from $t = 3$ to ensure a fair comparison. The corresponding VRMSE values are reported in \Cref{tab:vrmse_rollout}.

\begin{table*}[!ht]
    \centering
     \resizebox{\textwidth}{!}{\begin{tabular}{c|cccc}
    \hline
    Model & VRMSE($\rho$) & VRMSE($P$) & VRMSE($U$) & VRMSE($V$) \\
    \hline
    ConvNext U-Net\citep{ohana2024well} & 0.4127 & 6.0781 & 0.6235 & 4.2451 \\
    Baseline PARCv2 (\citep{nguyen2024parcv2} improved)& \textbf{0.4002} & 2.9766 & \textbf{0.4637} & 0.9583 \\
    MRPARCv2 (This work) & 0.4224 & \textbf{1.6466} & 0.4775 & \textbf{0.9412} \\
    \hline
    \end{tabular}}
    \caption{Roll-out prediction VRMSE comparison. Model with smallest VRMSE are in bold for each channel.}
    \label{tab:vrmse_rollout}
\end{table*}

We observe that errors accumulate rapidly during roll-out, with VRMSE values reaching 5–10 times those observed in single-step predictions. This rapid error growth can be partially attributed to the chaotic nature of turbulent flow, in which small deviations in the early stages of a simulation can lead to large discrepancies in later stages. When comparing MRPARCv2 to the baseline PARCv2, we find a notable $\sim$50\% reduction in pressure VRMSE. Given that pressure is the most challenging field to predict, as established in the previous subsection, this result highlights a significant and encouraging improvement enabled by the multi-resolution architecture.

On the other hand, the VRMSE for Y velocity shows only a modest reduction of $\sim$2\%, while the VRMSE for density and X velocity slightly increases by $\sim$3\% and $\sim$6\%, respectively. This divergent behavior warrants further investigation. Taking into account the chaotic nature of turbulent flow, the small amount of differences in earlier timesteps rapidly becomes large, therefore rendering roll-out VRMSE less sensitive to accuracy improvements observed in single-step predictions. In the next section, we provide a visual inspection of the predicted sequences to better understand the qualitative differences in model performance.

\subsection{Visual inspection of roll-out prediction}
\begin{figure}
    \centering
    \includegraphics[width=0.78\columnwidth]{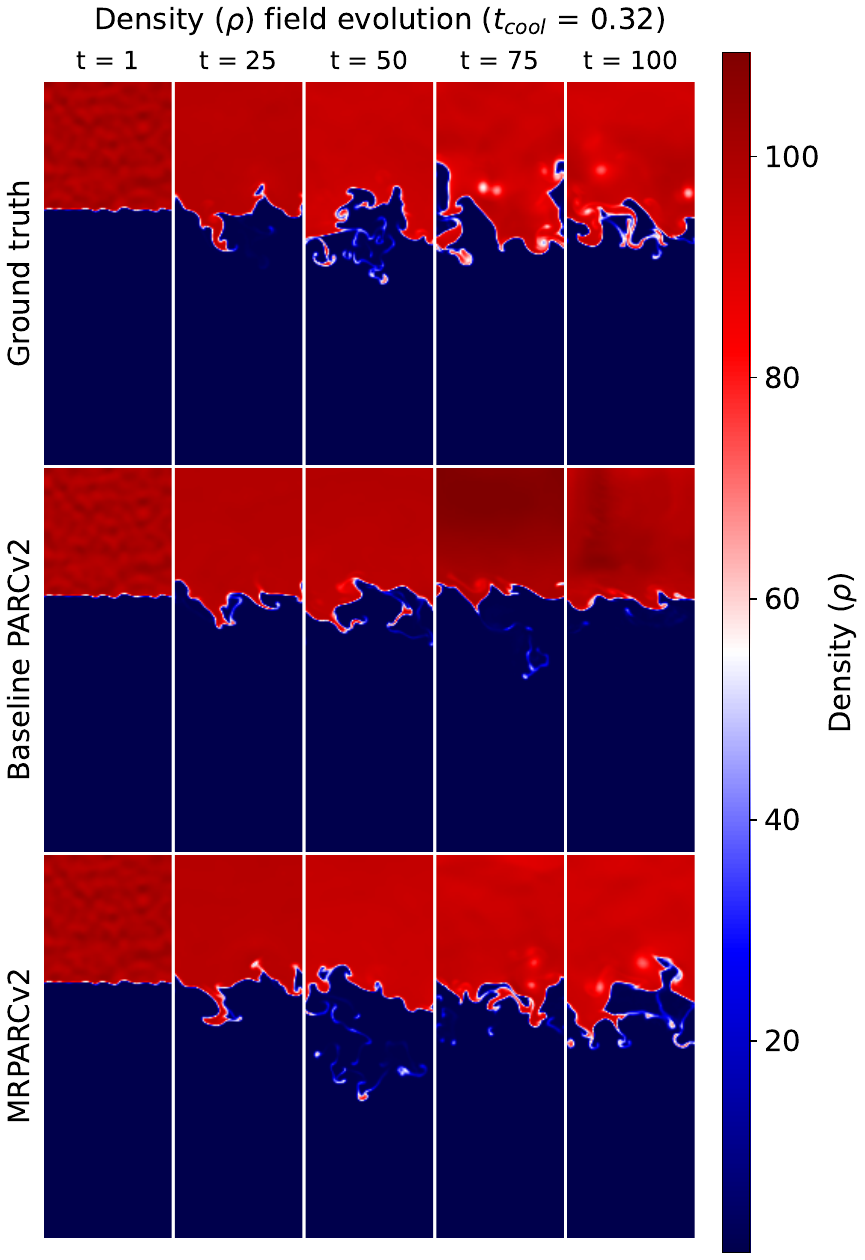}
    \caption{Ground truth and roll-out prediction of density field of $t_{cool}$ = 0.32}
    \label{fig:visual_rho_0.32}
\end{figure}

\begin{figure}
    \centering
    \includegraphics[width=0.78\columnwidth]{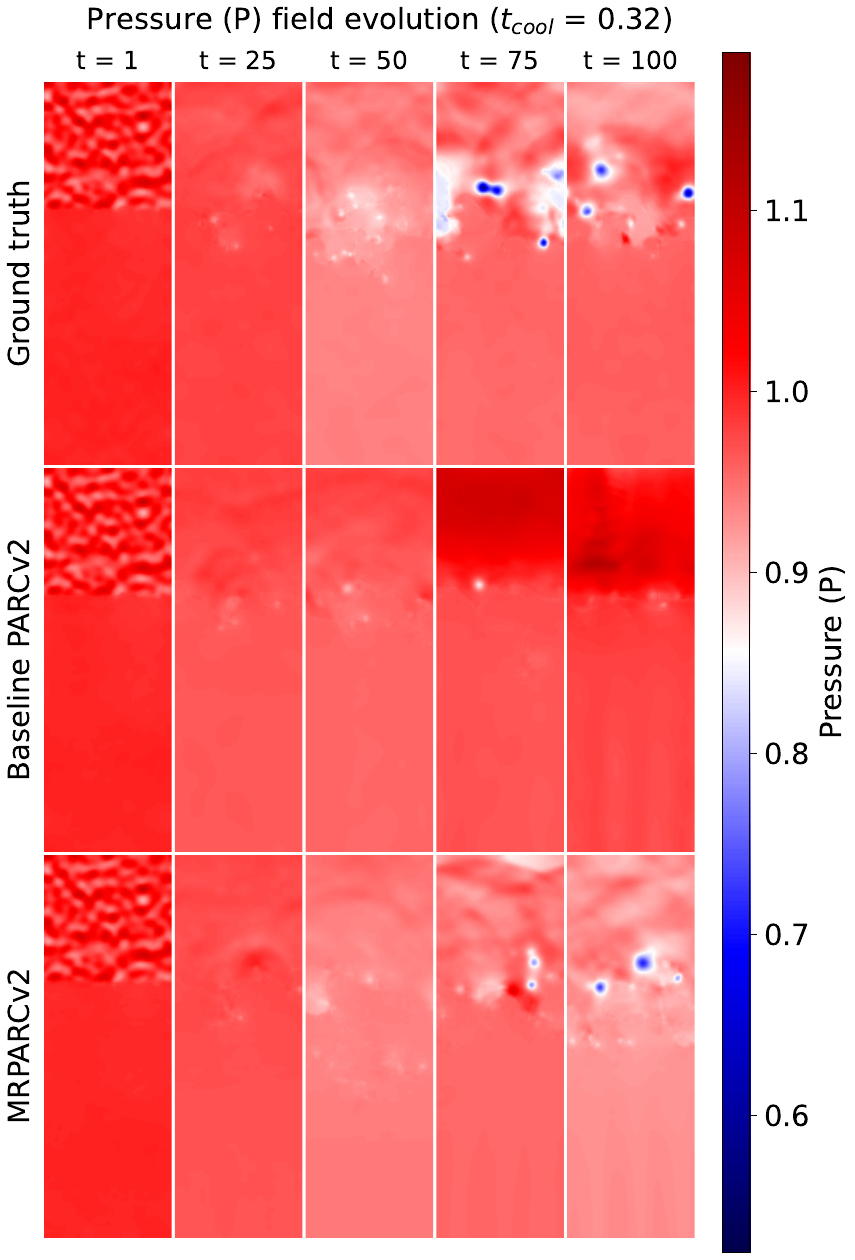}
    \caption{Ground truth and roll-out prediction of pressure field of $t_{cool}$ = 0.32}
    \label{fig:visual_p_0.32}
\end{figure}

\begin{figure}
    \centering
    \includegraphics[width=0.78\columnwidth]{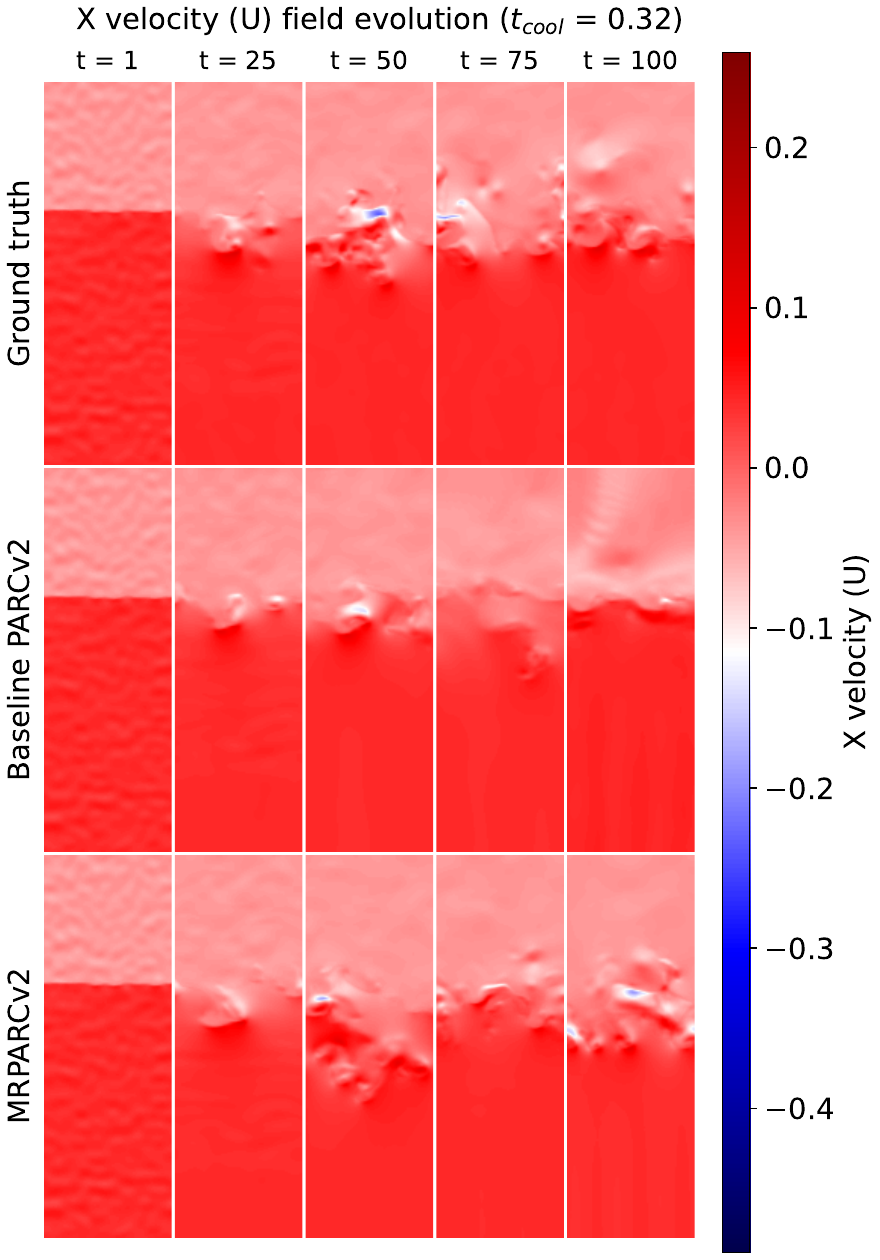}
    \caption{Ground truth and roll-out prediction of X velocity field of $t_{cool}$ = 0.32}
    \label{fig:visual_u_0.32}
\end{figure}

\begin{figure}
    \centering
    \includegraphics[width=0.78\columnwidth]{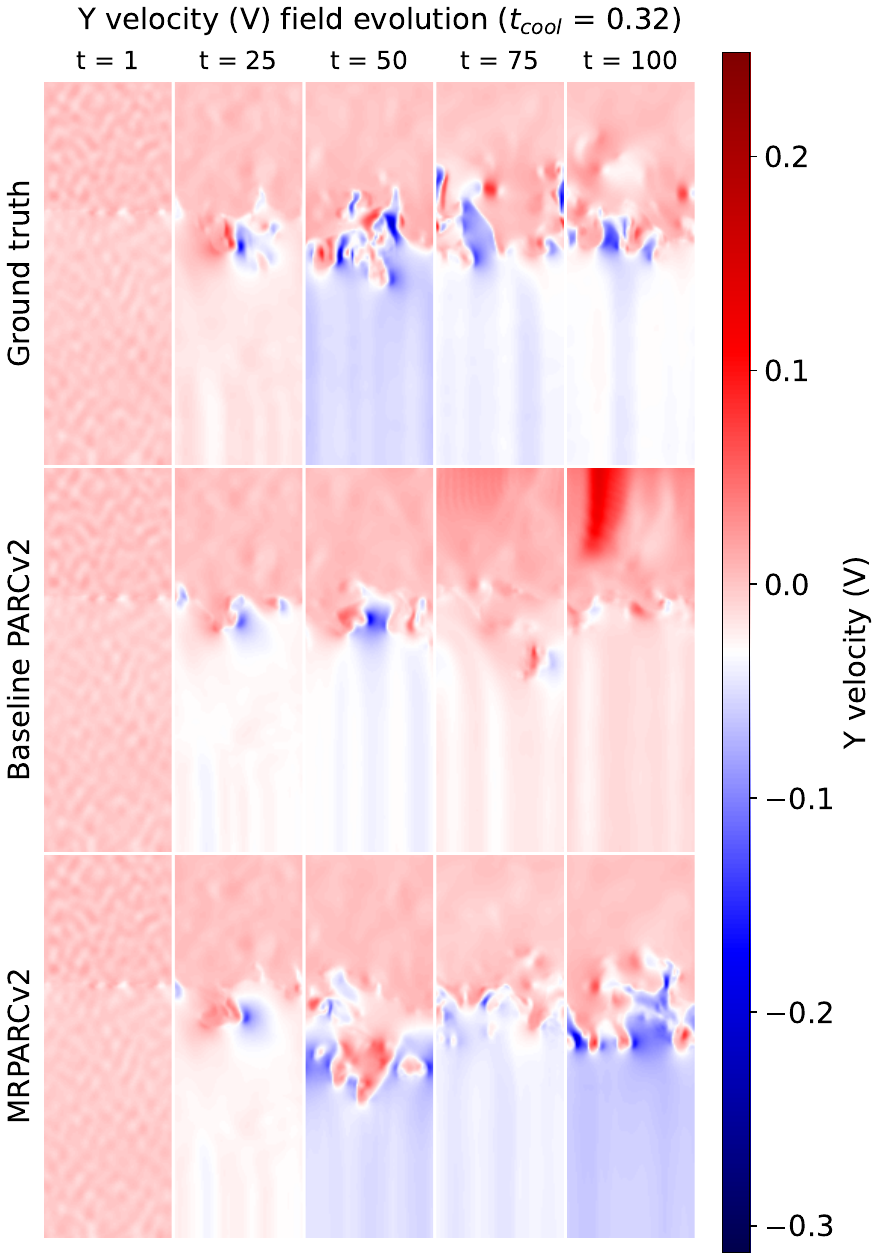}
    \caption{Ground truth and roll-out prediction of Y velocity field of $t_{cool}$ = 0.32}
    \label{fig:visual_v_0.32}
\end{figure}

We begin by evaluating model performance on the test case with $t_{cool} = 0.32$, which lies near the center of the training set distribution and exhibits the lowest prediction error among all test cases. Comparisons between the ground truth and the roll-out predictions from the baseline PARCv2 and MRPARCv2 models for the density, pressure, X velocity, and Y velocity fields are presented in \Cref{fig:visual_rho_0.32}--\Cref{fig:visual_v_0.32}, respectively.

Both models are able to predict the immediate next timestep with reasonable accuracy. Minimal differences are observed in the first column ($t=1$) of \Cref{fig:visual_rho_0.32}--\Cref{fig:visual_v_0.32}, regardless of the physical quantity being predicted. However, visible discrepancies between the model predictions begin to emerge as early as timestep $t = 25$, with MRPARCv2 showing greater fidelity than the baseline, particularly in capturing finer spatial details. In the density field, the baseline PARCv2 prediction exhibits reduced phase mixing compared to MRPARCv2, where the high-density phase penetrates much deeper into the low-density region at $t = 50, 75, 100$ in \Cref{fig:visual_rho_0.32}. The pressure field displays the most significant divergence between models. A clear distinction can be seen between the baseline PARCv2 and MRPARCv2 predictions: MRPARCv2 is able to accurately capture low-pressure regions, while the baseline model fails to do so and even overestimates pressure at $t = 75$ and $t = 100$, as shown in \Cref{fig:visual_p_0.32}. Similar patterns are observed in the X and Y velocity fields, as illustrated in \Cref{fig:visual_u_0.32} and \Cref{fig:visual_v_0.32}. Given the smaller number of trainable parameters in MRPARCv2, we attribute the improved performance of MRPARCv2 to the multi-resolution architecture. This enhancement is particularly valuable in modeling turbulent flow, as it improves the model’s capacity to represent finer-scale dynamics and smaller eddies—an essential component in accurately capturing turbulent behavior.

\begin{figure}
    \centering
    \includegraphics[width=0.78\columnwidth]{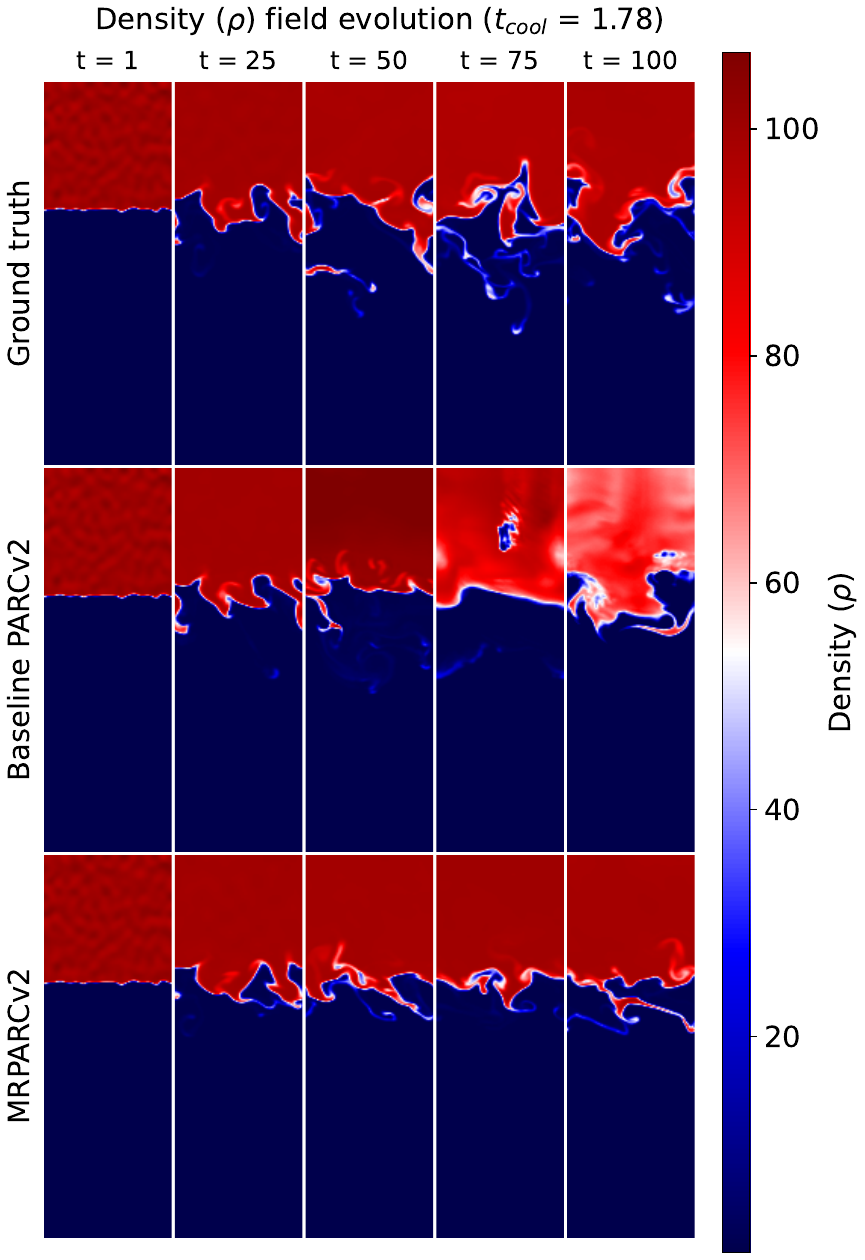}
    \caption{Ground truth and roll-out prediction of density field of $t_{cool}$ = 1.78}
    \label{fig:visual_rho_1.78}
\end{figure}
\begin{figure}
    \centering
    \includegraphics[width=0.78\columnwidth]{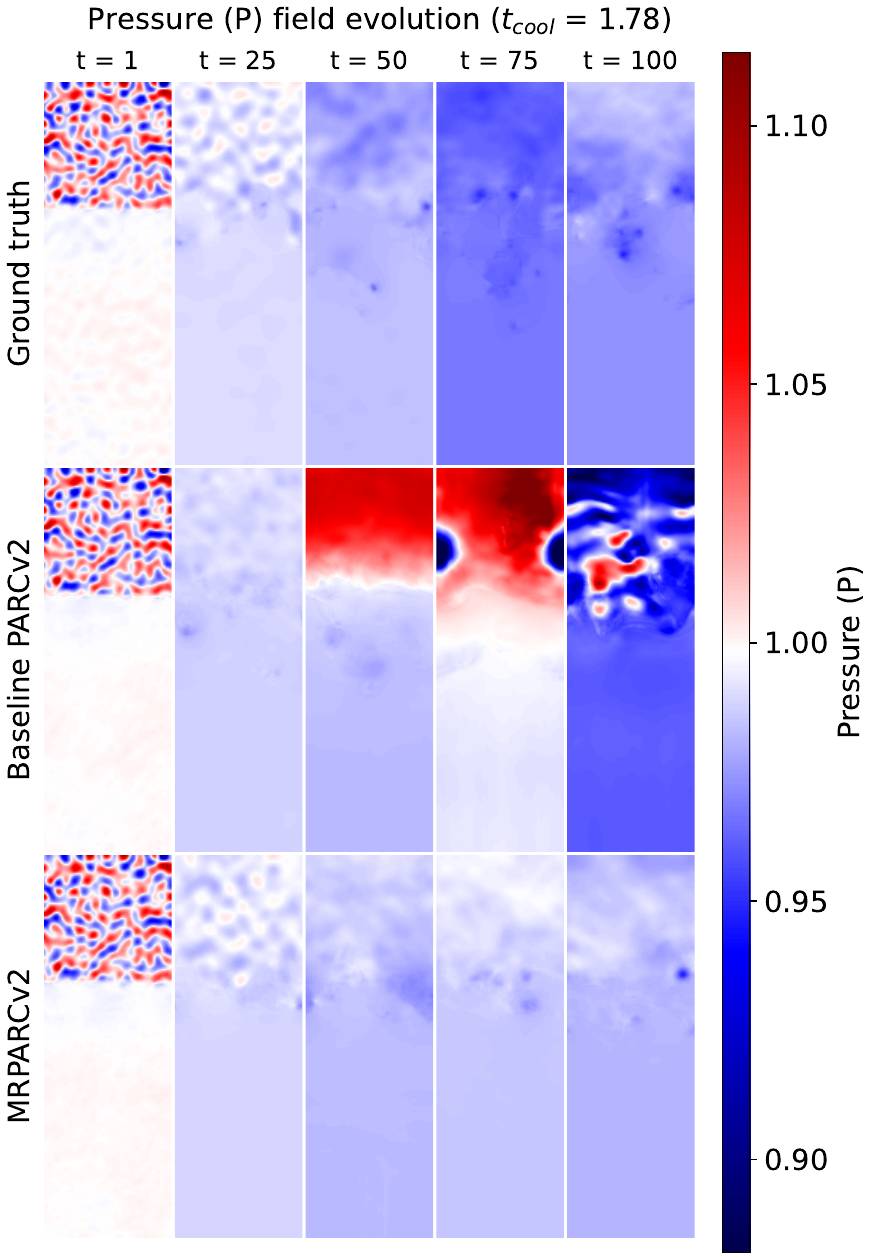}
    \caption{Ground truth and roll-out prediction of pressure field of $t_{cool}$ = 1.78}
    \label{fig:visual_p_1.78}
\end{figure}

We next examine the test case with $t_{cool} = 1.78$, which represents the slowest cooling system in our test set and exhibits the most complex pressure patterns. For brevity, we focus our discussion on the density and pressure fields. Numerical instabilities are immediately evident in the baseline model: large-scale deviations appear in the pressure field at $t = 50$ (second row of \Cref{fig:visual_p_1.78}) and in the density field at $t = 75$ (second row of \Cref{fig:visual_rho_1.78}). Even as early as $t = 25$, the baseline model fails to preserve high-pressure regions, indicating a breakdown in predictive accuracy. In contrast, MRPARCv2 maintains stability across both density and pressure fields throughout the prediction window (third rows of \Cref{fig:visual_p_1.78} and \Cref{fig:visual_rho_1.78}). Many high- and low-pressure cells are correctly captured, although there is a slight overestimation of the mean temperature field, as indicated by an overall lighter blue hue when compared to the ground truth.

\begin{figure}
    \centering
    \includegraphics[width=0.78\columnwidth]{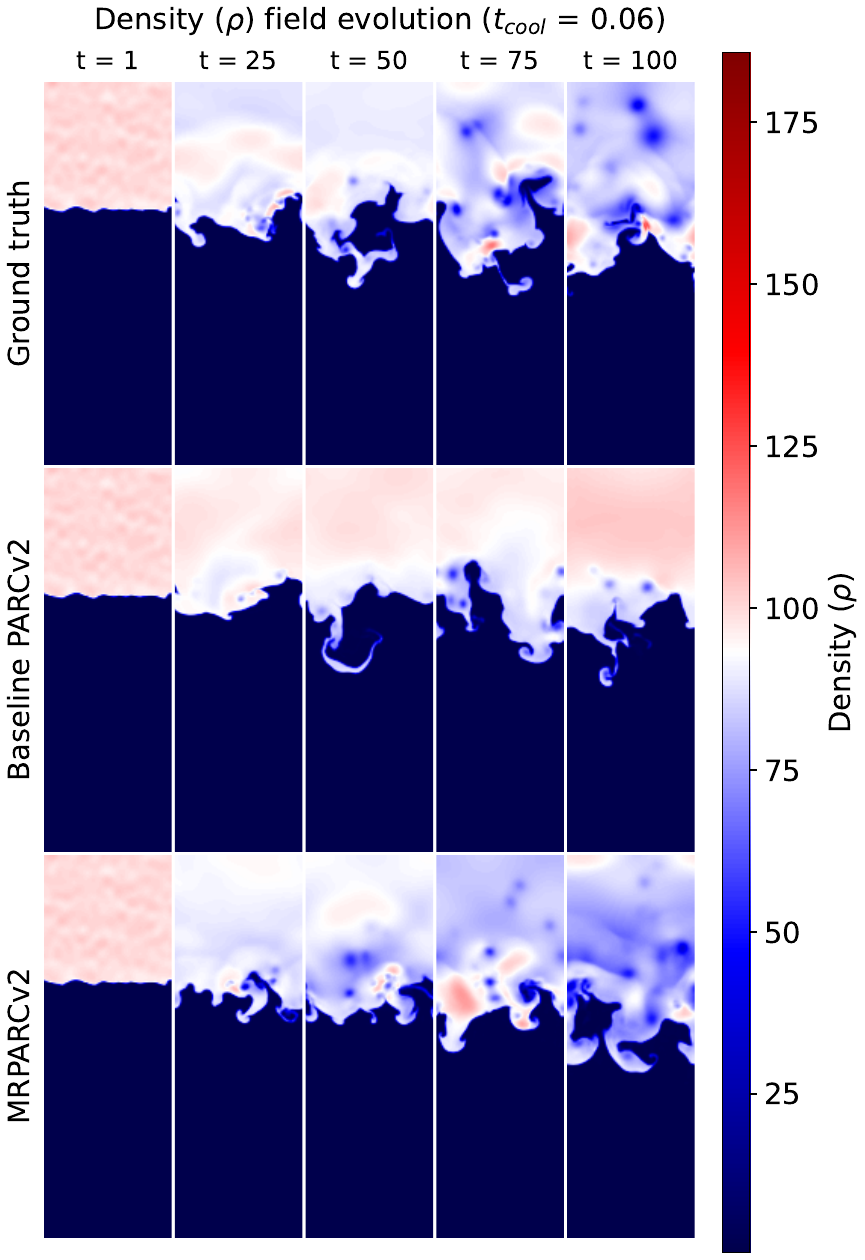}
    \caption{Ground truth and roll-out prediction of density field of $t_{cool}$ = 0.06}
    \label{fig:visual_rho_0.06}
\end{figure}
\begin{figure}
    \centering
    \includegraphics[width=0.78\columnwidth]{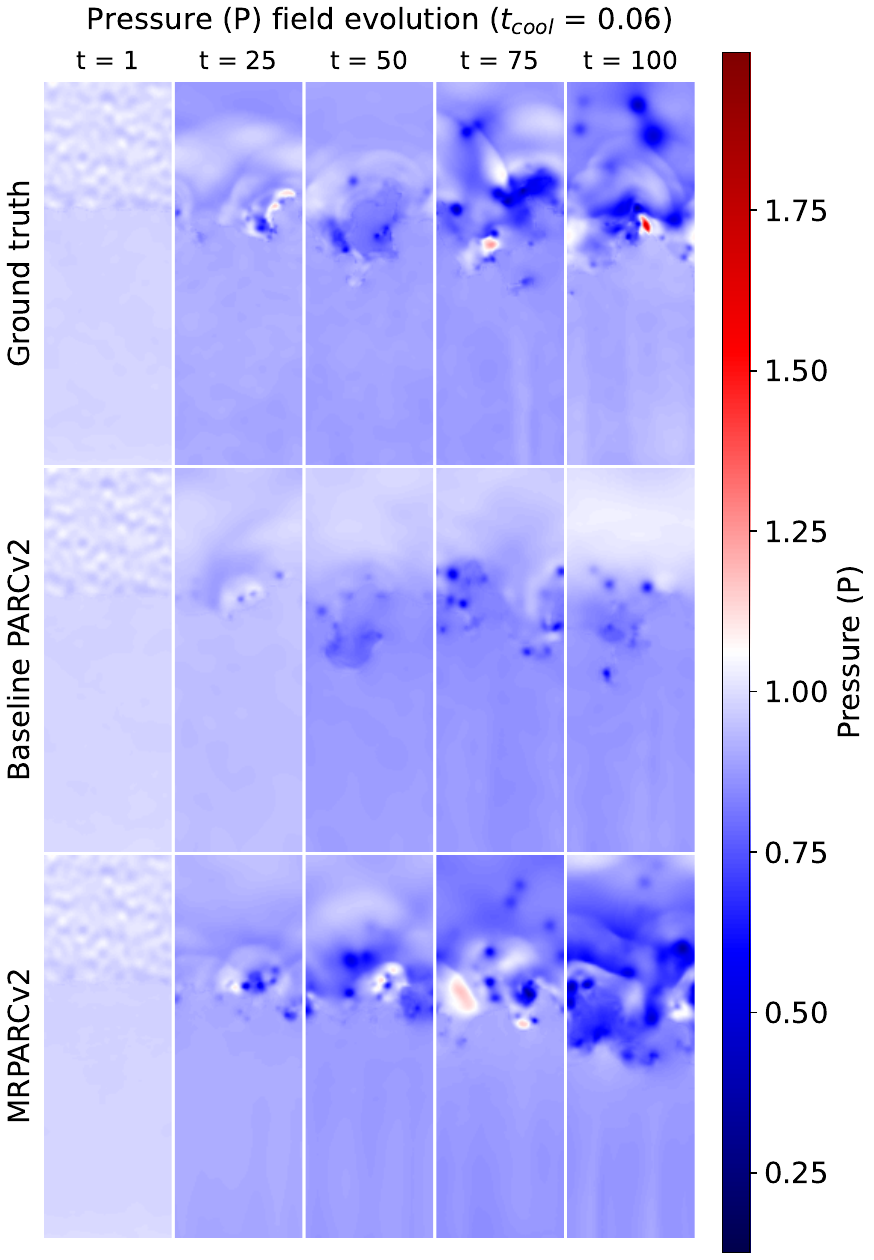}
    \caption{Ground truth and roll-out prediction of pressure field of $t_{cool}$ = 0.06}
    \label{fig:visual_p_0.06}
\end{figure}

Finally, we evaluate the test case with $t_{cool} = 0.06$, the fastest cooling system in our test set and the one with the most intricate density structures. MRPARCv2 accurately predicts strong mixing of the high-density phase, consistent with the ground truth in \Cref{fig:visual_rho_0.06}, where the high-density phase effectively ``disappears'' beyond $t \geq 25$. In the pressure field (\Cref{fig:visual_p_0.06}), the ground truth shows complex pressure front patterns for $t \geq 25$. These fronts remain relatively sharp in MRPARCv2’s predictions but are almost entirely smoothed out in those of the baseline model. Furthermore, both the ground truth and MRPARCv2 predictions display an expansion of low-pressure regions over time—a behavior that is much less pronounced in the baseline model.

Overall, we draw the following conclusions from the visual examination of the roll-out sequences. First, MRPARCv2 significantly improves the prediction of the pressure field, capturing both finer-scale structures (e.g., high- and low-pressure cells) and the overall mean field value, regardless of the $t_{cool}$ value. Second, MRPARCv2 exhibits more thorough phase mixing compared to the baseline model, resulting in predictions that are closer to the ground truth, particularly in fast cooling systems. Third, MRPARCv2 demonstrates greater numerical stability than the baseline model, especially in slow cooling regimes. We attribute all of these improvements to the adoption of the multi-resolution architecture.

\subsection{Physics-driven metrics}
VRMSE (or any other standard prediction error metric) suffers from a significant limitation when used to evaluate the performance of PIML models on complex dynamical systems. Due to the inherently sensitive nature of these systems, even small deviations in prediction can lead to rapid divergence from the ground truth, resulting in large error values over time. A similar phenomenon was observed in \citet{pfaff2020learning} when modeling the flag-blowing-in-wind problem, where predictions diverged significantly from the ground truth after only 50 time steps. \citet{ohana2024well} also emphasized the need for more physics-informed metrics to better evaluate the performance of machine learning models in such settings. Motivated by these insights, we selected the following two physics-driven metrics to compare MRPARCv2 against the baseline model.

\subsubsection{Turbulent kinetic energy (TKE) spectrum}
\begin{figure}
    \centering
    \includegraphics[width=0.91\columnwidth]{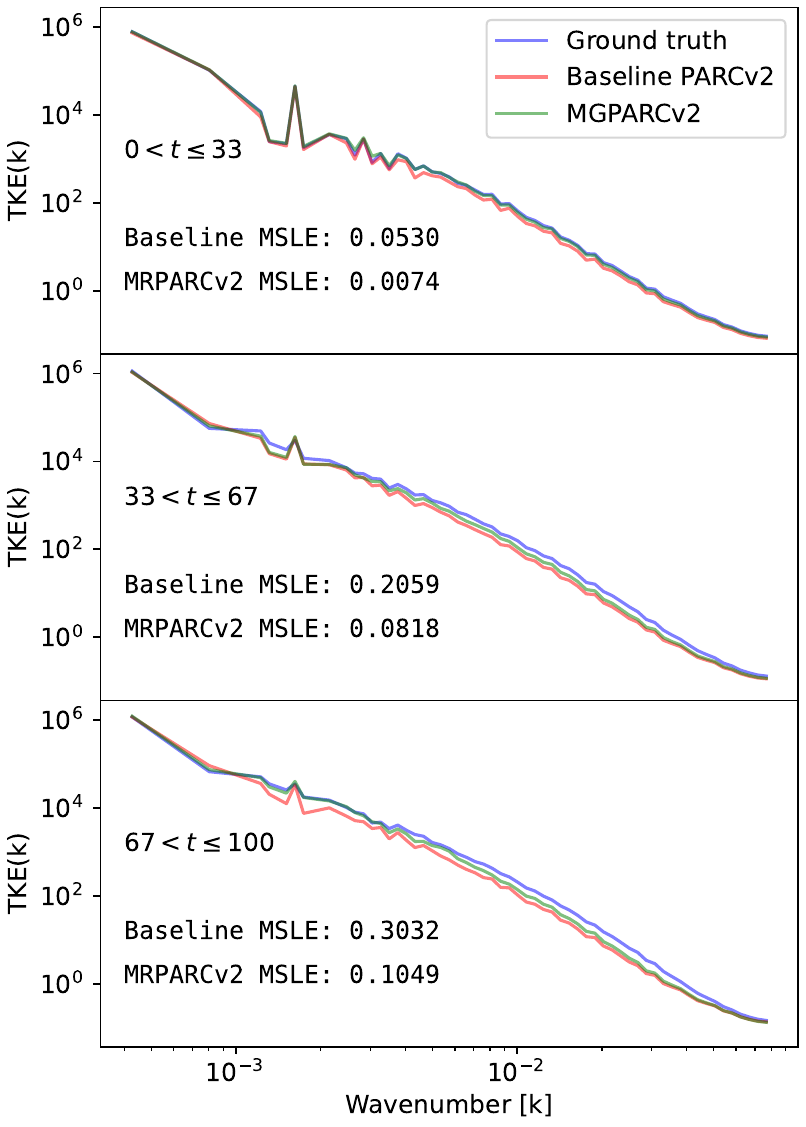}
    \caption{Turbulent kinetic energy spectrum for early ($0<t\leq 33$), middle ($33<t\leq 67$) and late stage ($67<t\leq 100$) of simulation.}
    \label{fig:pdm_tke_spectrum}
\end{figure}

Turbulent kinetic energy (TKE) represents the average kinetic energy per unit mass in a turbulent flow, and its spectral distribution has been widely used to validate CFD solvers (e.g.,\citep{robertson2015validation,saad2017scalable}). We therefore conducted a similar analysis for our PIML models. TKE spectra were computed for each simulation and then averaged over timesteps and $t_{cool}$ values to generate \Cref{fig:pdm_tke_spectrum}, which shows results at early (top panel), middle (middle panel), and late (bottom panel) stages of the simulations. While the TKE spectrum tends to be noisy in the higher frequency range, it generally exhibits limited variation over time and across different $t_{cool}$ values.

In addition, we report the mean squared logarithmic error (MSLE) for both the baseline model and MRPARCv2, computed using the following equation:
\begin{equation}
    \text{MSLE}(y, \hat{y}) = \frac{1}{N}\sum(\log(y_i)-\log(\hat{y}_i))^2
\end{equation}
where $y_i$ denotes the ground truth TKE spectrum and $\hat{y}_i$ represents the predicted spectrum. MSLE is preferred over mean squared error (MSE) in this context because the TKE spectrum spans several orders of magnitude; using MSE would disproportionately emphasize performance on the largest eddies, while MSLE provides a more balanced evaluation across scales.

We observe that while both models experience error accumulation—as indicated by increasing MSLE values and growing deviations from the ground truth spectrum in the later stages of the simulation—MRPARCv2 consistently outperforms the baseline model. Specifically, MRPARCv2 achieves a $\sim$86\% reduction in MSLE during the early stage and a $\sim$67\% reduction during the late stage of the simulation. In fact, during the early stage, the MRPARCv2 TKE spectrum is nearly indistinguishable from the ground truth, providing strong evidence that MRPARCv2 captures the underlying governing equations more effectively than the baseline PARCv2. This result further highlights the benefit of incorporating the multi-resolution architecture. An interesting observation is that both models tend to under-predict the total kinetic energy of the fluid, as indicated by the red and green curves consistently falling below the ground truth (blue) curve. This suggests that inductive bias models may exhibit more diffusive rather than oscillatory behavior, a hypothesis that merits further investigation.

\subsubsection{Mass-temperature distribution}
\begin{figure}
    \centering
    \includegraphics[width=0.91\columnwidth]{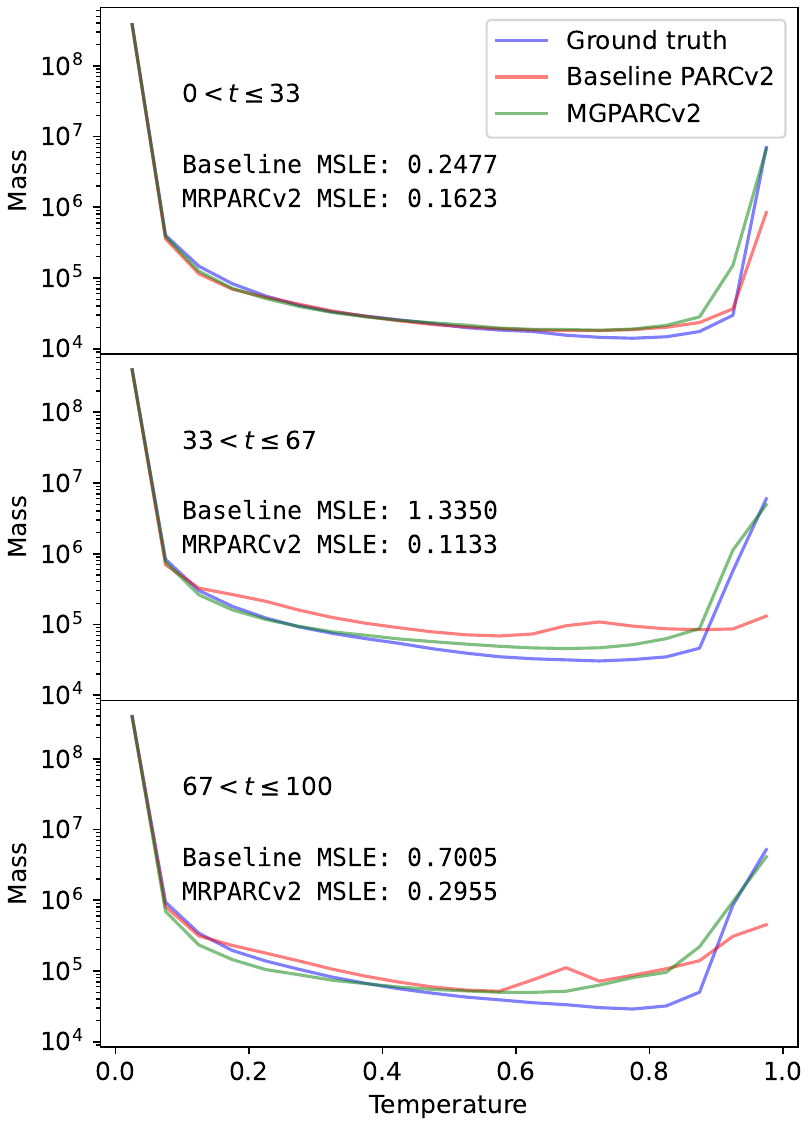}
    \caption{Mass-temperature relation for early ($0<t\leq 33$), middle ($33<t\leq 67$) and late stage ($67<t\leq 100$) of simulation.}
    \label{fig:pdm_mass_temperature}
\end{figure}

As one of the key driving physical processes in this system is phase mixing, a useful physics-driven metric is the mass–temperature distribution. This metric is computed by first deriving temperature from the fields using \Cref{eqn:eos}, then binning the results by temperature and summing the corresponding density values within each bin. The mass–temperature distribution is computed for each simulation and then averaged over timesteps and $t_{cool}$ values to produce \Cref{fig:pdm_mass_temperature}, which shows results at the early (top panel), middle (middle panel), and late (bottom panel) stages of the simulations. As with the TKE analysis, mean squared logarithmic error (MSLE) is used to quantitatively evaluate model performance.

We observe that, once again, MRPARCv2 significantly outperforms the baseline PARCv2 model, particularly in the high-temperature phase. Notable deviations in the baseline model begin as early as $t \leq 33$, and its predictions progressively under-predict the mass in the hot phase as the simulation proceeds. In the middle stage of the simulation, the hot phase nearly disappears entirely from the baseline model’s prediction. In contrast, MRPARCv2 yields a mass–temperature distribution that more closely resembles the ground truth, with MSLE reductions ranging from 35\% to 92\% depending on the simulation stage.

However, visual inspection of the distribution curves reveals that both models still exhibit substantial deviations from the ground truth. We attribute this discrepancy to the models learning the equation of state less accurately than the conservation laws. In the following section, we provide a detailed analysis to support this attribution and discuss its implications.

\section{Discussion}\label{sec:discussion}
\subsection{Multi-resolution Feature Ablation Study}
\begin{figure}
    \centering
    \includegraphics[width=0.73\columnwidth]{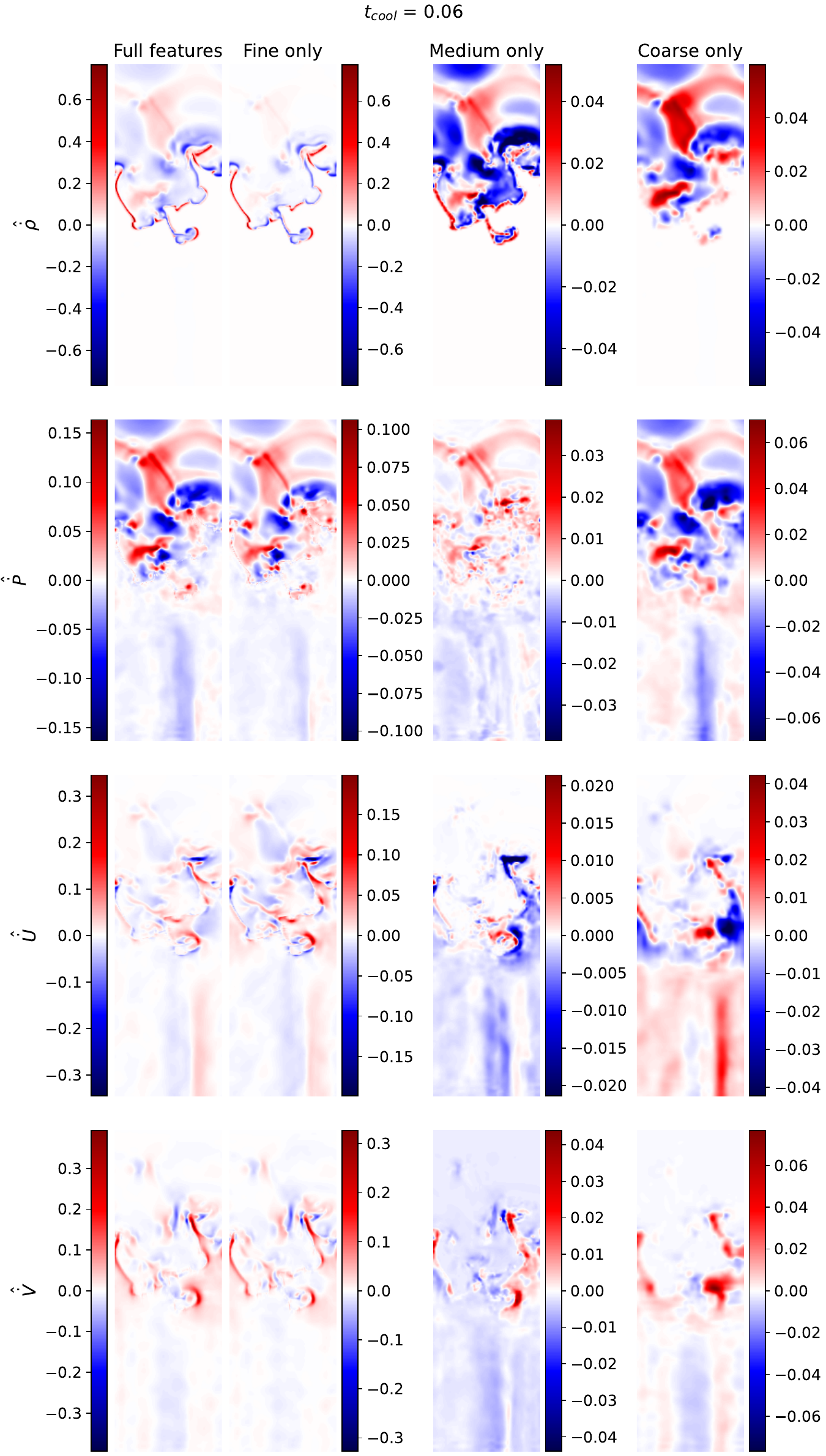}
    \caption{Model predicted temporal derivatives with features of different resolution for $t_{cool}$ = 0.06 at t = 75.}
    \label{fig:mr_ablation_0.06}
\end{figure}

\begin{figure}
    \centering
    \includegraphics[width=0.73\columnwidth]{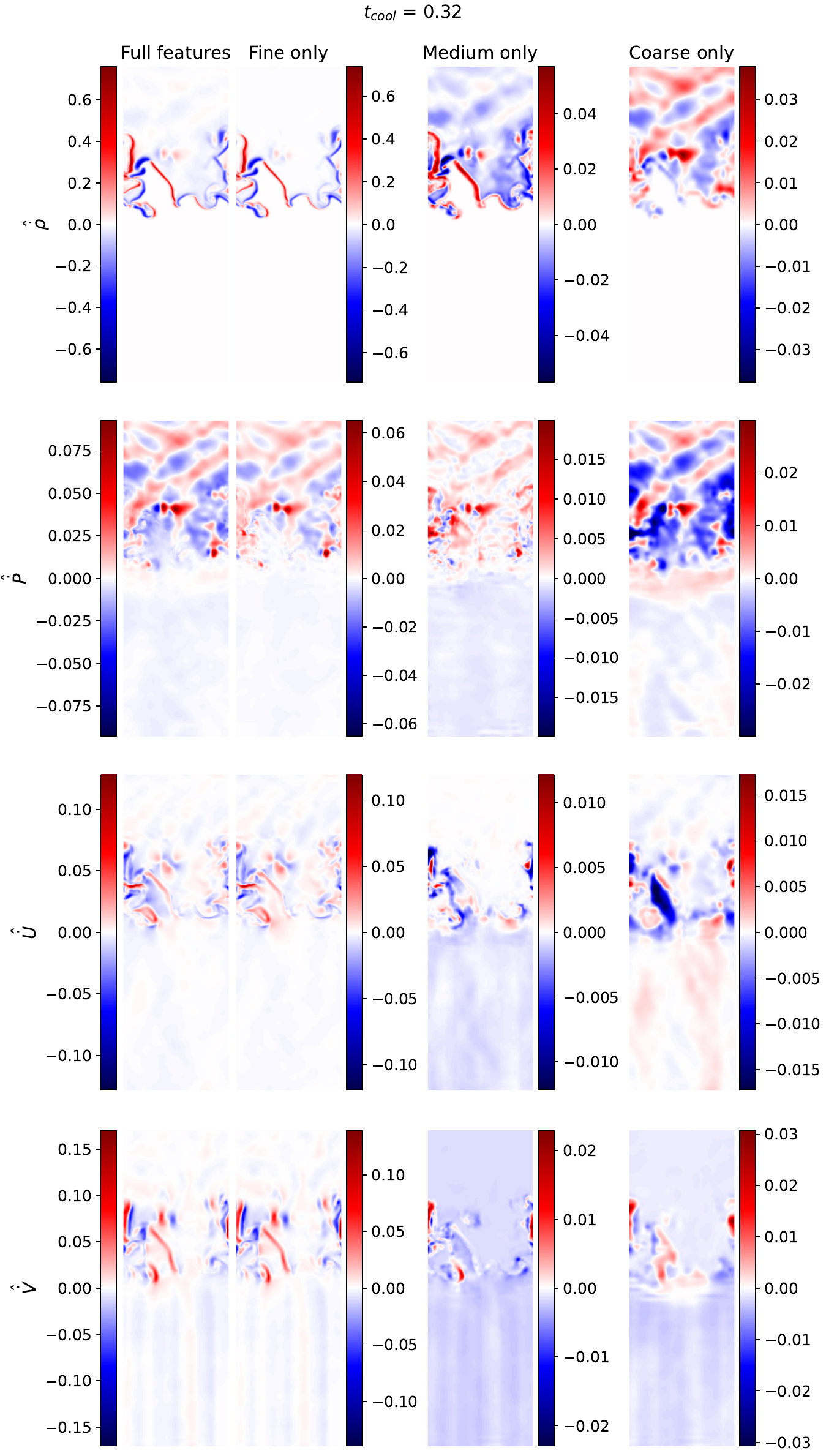}
    \caption{Model predicted temporal derivatives with features of different resolution for $t_{cool}$ = 0.32 at t = 75.}
    \label{fig:mr_ablation_0.32}
\end{figure}

\begin{figure}
    \centering
    \includegraphics[width=0.73\columnwidth]{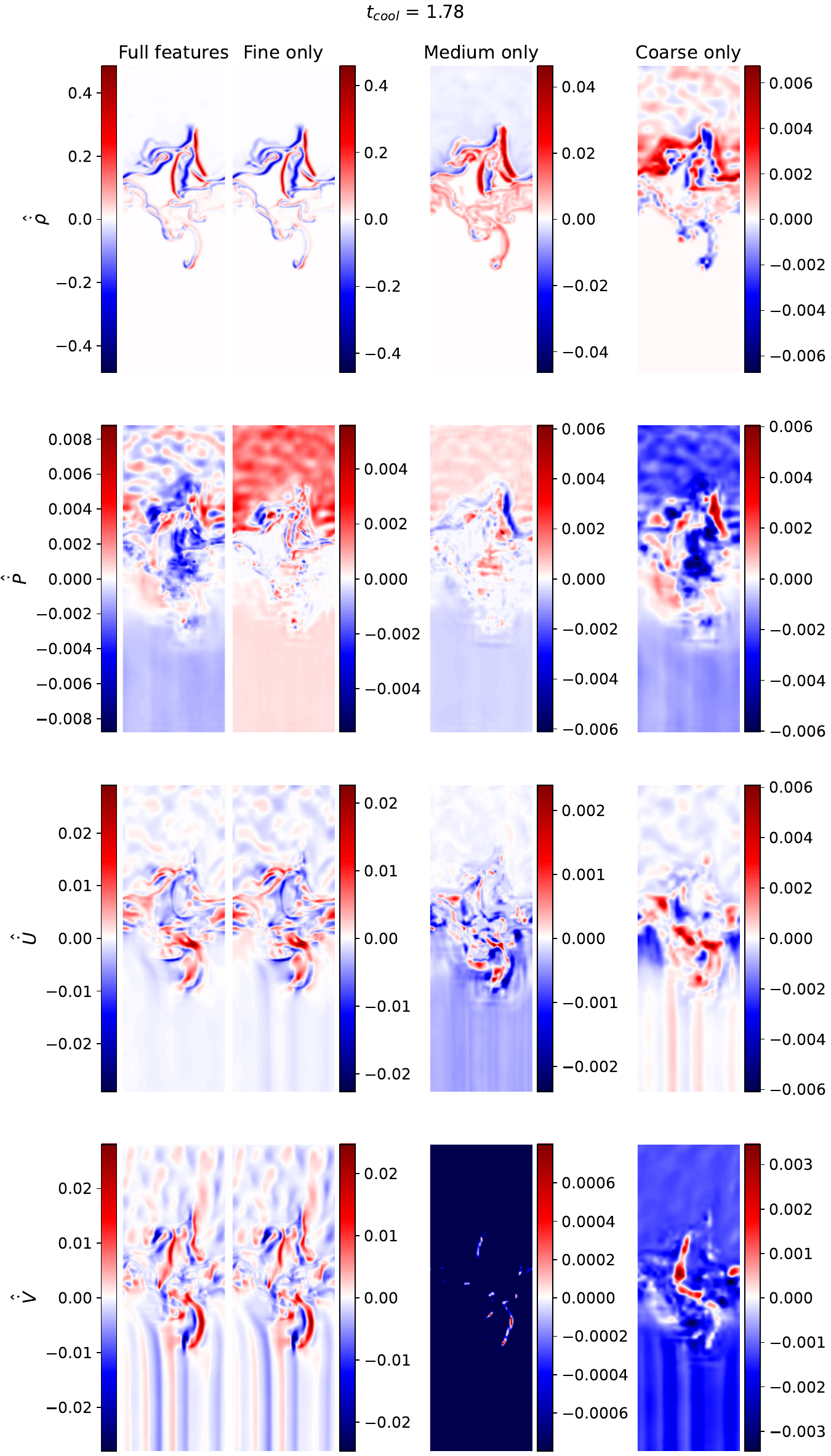}
    \caption{Model predicted temporal derivatives with features of different resolution for $t_{cool}$ = 1.78 at t = 75.}
    \label{fig:mr_ablation_1.78}
\end{figure}

We assess the role of each resolution in predicting time derivatives via an ablation study, retaining features from only one resolution while zeroing out the others. Predictions from each single-resolution setting are compared with the full three-resolution model (\Cref{fig:mr_ablation_0.06}-\Cref{fig:mr_ablation_1.78}). The analysis focuses on $t=75$, when flow structures are most complex. Resolution contributions depend on both the gas cooling time ($t_{cool}$) and the physical quantity predicted. The clearest ``division of labor” occurs for density (first row of \Cref{fig:mr_ablation_0.06}-\Cref{fig:mr_ablation_1.78}): fine-resolution features capture phase-interface changes, while coarse-resolution features represent large-scale density variations within the high-density phase, which arises from the cooling of hot gas in cold phase. High rate-of-change regions typically occupy small spatial fractions, explaining the larger magnitudes from fine-resolution predictions. These findings highlight the natural suitability of multi-resolution architectures for modeling complex, multi-scale flows.

For pressure, we observed a markedly different trend. Features from the coarser resolution contribute substantially more: the maximum magnitude of the predicted time derivative from the coarse resolution alone is between $\sim$50\% and $\sim$75\% of the maximum from all three resolutions, increasing with larger $t_{cool}$ values. In contrast, the corresponding ratio for density is only around 5\%. A complete breakdown is observed for $t_{cool}=1.78$ in the second row of \Cref{fig:mr_ablation_1.78}, where predictions from the finest-resolution features alone significantly over-predict the pressure field and fail to align with the prediction from all three levels of features. This behavior likely explains the numerical failure observed in the baseline single-resolution model in \Cref{fig:visual_p_1.78}, as well as the large improvement in pressure accuracy during roll-out. From a machine learning perspective, the ground-truth pressure fields exhibit distinct patterns of localized low-pressure cells surrounded by higher-pressure regions. In contrast, the density and velocity fields show large continuous regions of nearly uniform values in the hot and cold phases, with rapid variations confined primarily to the interface. We hypothesize that pressure-like patterns are inherently harder to learn, as spectral bias in neural networks has a stronger impact on prediction accuracy, beyond simply smearing out the phase interface. From a computational physics perspective, solving the pressure-velocity coupling typically requires multiple iterations of the pressure-implicit with splitting of operators (PISO) algorithm to achieve convergence, which poses an additional challenge for a single-resolution baseline model to capture. While we do not observe a significant reduction in velocity prediction errors, we do find a substantial decrease in the error of the turbulent kinetic energy spectrum.

The behavior of the velocity field appears more difficult to interpret. We do not observe a clear separation of roles as seen in density, nor a dramatic failure as in the finest-resolution-only pressure predictions. These findings suggest that encouraging a separation of roles across all four fields could be a promising direction for achieving further improved prediction accuracy.

\subsection{Learning the Equation of State}
As previously noted, the error in the mass–temperature distribution is considerably more pronounced than in other physics-driven metrics or VRMSE values. This observation suggests a potentially larger prediction error in the derived temperature field. Since temperature is neither provided as an input nor as an output channel in the dataset, the neural network has not been explicitly exposed to this physical quantity during training, making it inherently more difficult to predict accurately. Furthermore, as an inductive bias model, the absence of an embedded equation of state within the PARCv2 architecture, both baseline and multi-resolution variants, further limits its capacity to reconstruct temperature correctly. Additionally, \citet{liu2024kan} noted that division operations, such as those involved in computing temperature, pose a particular challenge for neural networks using ReLU-family activation functions.

To further investigate this issue, we examine the VRMSE and distribution of the derived temperature. Upon inspection, we observed a small number of extreme outliers in the predicted temperature values from both models, several predictions are 5-6 orders of magnitude greater than the mean. Including these outliers in statistical analyses or visualizations would obscure the actual performance on the majority of the predictions and distort any meaningful interpretation of model accuracy.

To mitigate this, we clip the predicted temperature values to lie within the range $[-10,10]$. This approach serves two purposes: (a) it makes visible any outliers that exceed the expected distribution based on the ground truth, and (b) it reduces their impact on descriptive statistics. It is worth noting that ground truth temperature values almost always lie within the range $[0,1]$. We present the VRMSE of the clipped temperature predictions for each $t_{cool}$ in \Cref{tab:vrmse_temperature}. This analysis is conducted on roll-out predictions and should therefore be compared directly to the VRMSE values reported in \Cref{tab:vrmse_rollout}.

\begin{table*}[!ht]
    \centering
     \resizebox{\textwidth}{!}{\begin{tabular}{c|cccccccc}
    \hline
    $t_{cool}$ & Average & 0.06 & 0.10 & 0.18 & 0.32 & 0.56 & 1.00 & 1.78 \\
    \hline
    Baseline PARCv2 & 0.8787 & 1.2285 & 0.9626 & 0.7977 & 0.7121 & 0.6131 & 0.4516 & 1.3853 \\
    MRPARCv2 & 0.5651 & 0.7292 & 0.7612 & 0.6406 & 0.5070 & 0.4476 & 0.4104 & 0.4600 \\
    \hline
    \end{tabular}}
    \caption{Roll-out prediction VRMSE comparison on temperature.}
    \label{tab:vrmse_temperature}
\end{table*}

As expected, MRPARCv2 outperforms the baseline model, achieving an average VRMSE reduction of approximately 36\%. When comparing across physical quantities, the derived temperature field demonstrates better accuracy than pressure and Y velocity, though this observation holds only when extreme outliers are clipped from the predictions. One potential avenue for future improvement is the adoption of Kolmogorov-Arnold Networks (KAN), which explicitly address the mathematical difficulty of learning division operations; however, exploring this approach is beyond the scope of the current work.

\subsection{Change of Variable (CoV)}
To address the shortcomings in temperature predictions noted above, caused by only four out of the five physical quantities being input during training, we explored an alternative approach to mitigate the issue of outliers and nonphysical predictions by changing the predicted variable from pressure to temperature. Given that both the baseline and MRPARCv2 models exhibit similar behavior in the derived temperature field, we conducted this experiment using only the baseline model. In this configuration, temperature is directly predicted, and pressure is subsequently derived using the equation of state. The average roll-out VRMSE across all five physical quantities, four predicted directly and one derived, is presented in \Cref{tab:vrmse_cov}.

\begin{table}[!ht]
    \centering
    \begin{tabular}{c|cc}
    \hline
     & VRMSE($T$) & VRMSE($P$) \\
    \hline
    Baseline PARCv2 & \underline{0.8787} & 2.9766 \\
    Baseline CoV & \textbf{0.4964} & \underline{7.9387} \\
    MRPARCv2 & \underline{0.5651} & \textbf{1.6466} \\
    \hline
    \end{tabular}
    \caption{Roll-out prediction VRMSE comparison on temperature and pressure, including the baseline change of variable (Baseline CoV) model. EOS-derived physical quantities for each model are underlined. Model with smallest VRMSE are in bold for each channel.}
    \label{tab:vrmse_cov}
\end{table}

As expected, the VRMSE of temperature is the lowest in the baseline CoV model, since the network is now explicitly trained to predict this variable. However, this improvement comes at the cost of increased VRMSE in pressure. Because pressure is computed as the product of temperature and density as in \Cref{eqn:eos}, this experiment challenges one of the previously proposed explanations that the reduced performance is due to the difficulty ReLU-family neural networks have in learning division operations. Instead, we recommend that future PIML modeling efforts for compressible flows include all physical quantities appearing in the governing equations as both input and output channels, even if the equation of state or closure models render one of them redundant.

We further hypothesize that the performance degradation in the EOS-derived quantity is largely due to the absence of inductive bias associated with the equation of state itself. Combined with the limited availability of training data, a common constraint in many AI-for-science applications, learning the EOS purely from data becomes an especially difficult task. While clipping the predictions makes the EOS-derived variable usable, a more robust solution would involve embedding the equation of state directly into the neural network architecture. However, this remains a challenging endeavor due to the often complex and problem-specific structure of EOS formulations, which makes the development of a universal method for EOS integration into neural networks an open and unsolved problem.

\subsection{Uncertainty Quantification}\label{sec:uq}
Uncertainty quantification (UQ) of deep learning models is an active area of research in both the computer vision \citep{mucsanyi2024benchmarking} and physics-informed machine learning \citep{sun2020physics,morimoto2022assessments} communities. In traditional machine learning, uncertainty is often assessed using techniques such as 
k-fold cross-validation. However, this approach requires retraining the model on multiple train–validation splits, which becomes computationally prohibitive for deep neural networks. To address this challenge, a variety of alternative methods have been developed to estimate aleatoric and/or epistemic uncertainty with minimal additional cost. Some neural architectures, such as probabilistic neural networks, are designed to output uncertainty directly as part of the prediction \citep{maulik2020probabilistic}. Gaussian process–based approaches, including the Laplace approximation \citep{daxberger2021laplace} and stochastic weight averaging Gaussian (SWAG) \citep{maddox2019simple}, also provide uncertainty estimates. In addition, ensemble-based strategies such as Monte Carlo dropout \citep{gal2016dropout,srivastava2014dropout} and deep ensembles \citep{lakshminarayanan2017simple} have emerged as effective and widely used techniques.

\begin{figure*}
    \centering
    \includegraphics[width=0.89\linewidth]{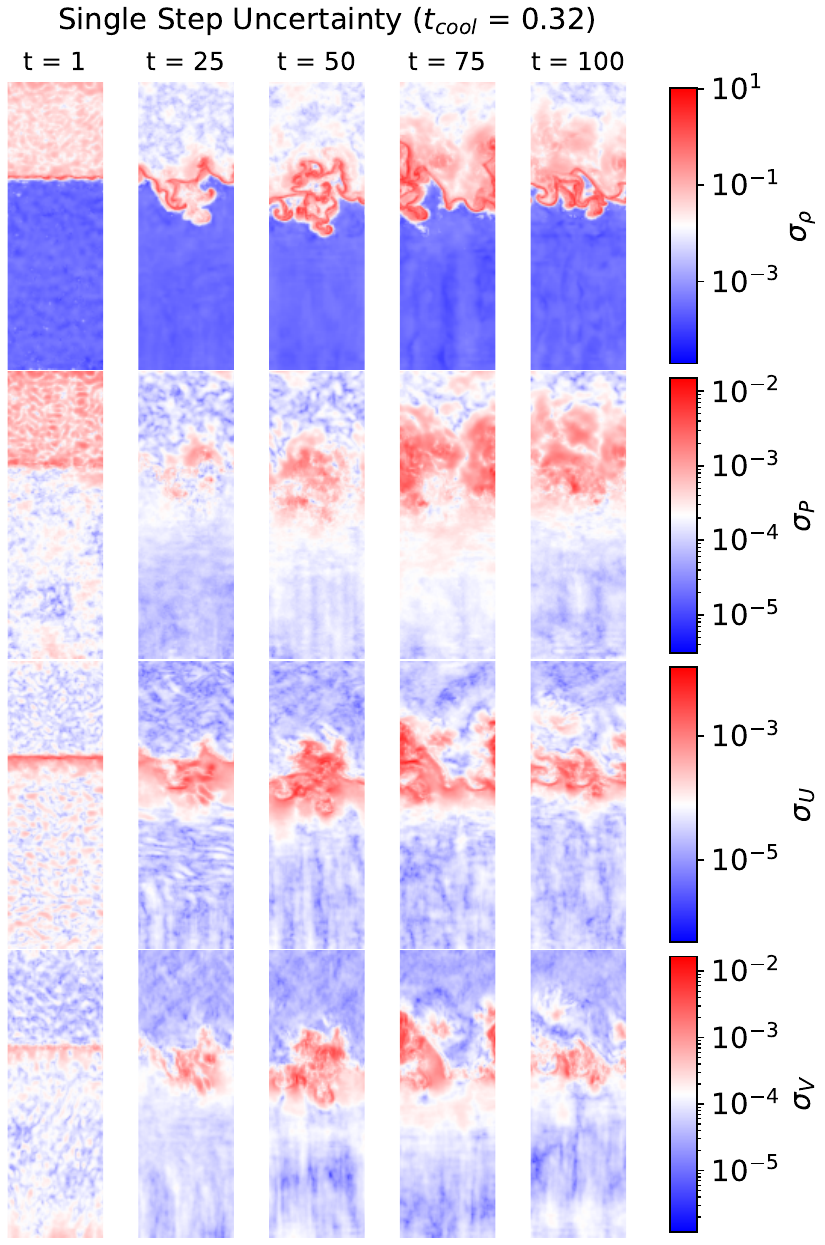}
    \caption{Epistemic uncertainty from MRPARCv2 for $t_{cool}=0.32$.}
    \label{fig:uq_single}
\end{figure*}

Here, we performed a pilot study to estimate the epistemic uncertainty of MRPARCv2 on a single test case. The uncertainty was quantified by performing inference with network weights from checkpoints within 
$\pm$15 epochs of the best validation loss, and the standard deviation across predictions was treated as the epistemic uncertainty. The result is presented in \Cref{fig:uq_single}. Across physical fields and simulation timesteps, the largest uncertainty consistently appears at the interface between the cold and hot phases, particularly in the later stages when turbulent eddies are more developed. Interestingly, we also observe a systematically smaller uncertainty in the hot phase for the density and pressure fields, whereas such behavior is absent in the velocity fields. As highlighted by \citet{mucsanyi2024benchmarking}, many current approaches struggle to accurately capture aleatoric uncertainty, and the most effective methods tend to be both task- and problem-specific. We are actively investigating different UQ techniques and their effectiveness for spatial–temporal regression problems.

\section{Conclusion}\label{sec:conclusion}
In this work, we proposed the multi-resolution Physics-Aware Recurrent Convolutional Neural Network (MRPARCv2), which integrates the structural embedding of the advection-diffusion-reaction (ADR) equations with a multi-resolution numerical approach. We demonstrated its superior accuracy in modeling turbulent flow, with significant error reductions observed across a variety of benchmark metrics, including both VRMSE-based prediction errors and physics-driven metrics. Given the 30\% smaller number of trainable parameters in MRPARCv2 model, the improvements can be largely attributed to the introduction of the multi-resolution architecture. We believe these findings are generalizable to other complex flow problems, as many complex flow problems are inherently multi-resolutional, and thus aligns well with the inductive bias introduced by a multi-resolution architecture.

We further investigated the model's capacity to learn the equation of state (EOS), a constraint not explicitly embedded in the network, and observed degraded performance in EOS-derived quantities. Our analysis showed that a simple change of variable does not resolve this issue; instead, it merely transfers the challenge to the new implicit variable. Improving the ability of PIML models to incorporate EOS constraints would provide substantial gains in accuracy for complex fluid simulations and broaden the applicability of PIML in computational physics. Future work should explore principled approaches to embedding EOS knowledge into neural architectures, despite the challenge posed by the diversity and complexity of EOS formulations across physical systems.

\section*{Acknowledgment}
This work was supported by the National Science Foundation under Grant No. DMREF-2203580.

\section*{Author Declarations}
The authors have no conflicts to disclose.

\section*{Data Availability}
The data that support the findings of this study are openly available in The Well multi-physics dataset, available at \url{https://polymathic-ai.org/the_well/}.

%\section*{Author Contributions}
%\textbf{XC:} Conceptualization, Data curation, Formal analysis, Methodology, Software, Visualization, Writing -- original draft, Writing –- review and editing; \textbf{JC:} Data curation, Validation, Visualization, Writing –- review and editing; \textbf{HU:} Funding acquisition, Project administration, Supervision, Writing –- review and editing; \textbf{SB:} Funding acquisition, Project administration, Resources, Supervision, Writing –- review and editing

\bibliography{aiptemplate}

\end{document}